\documentclass[article]{aa} %

\usepackage{txfonts}
\usepackage[utf8]{inputenc}
\usepackage{natbib}
\bibpunct{(}{)}{;}{a}{}{,} 
\usepackage{caption}
\usepackage{color}
\usepackage{subcaption}
\usepackage{float}
\usepackage{wasysym}
\usepackage{array}
\usepackage{textcomp}

\usepackage{txfonts}
\begin{document}
\title{Nebular dead zone effects on the D/H ratio in chondrites and comets}
\authorrunning{Ali-Dib et al.}

\author{M.~Ali-Dib\inst{1}, R. G.~Martin \inst{2}, J.-M.~Petit\inst{1}, O.~Mousis\inst{3}, P.~Vernazza\inst{3} and J.~I.~Lunine\inst{4}}
\institute{Institut UTINAM, CNRS-UMR 6213, Observatoire de Besan\c con, Universit\'{e} de Franche-Comt\'{e}, BP 1615, 25010 Besan\c{c}on Cedex, France
\email{mdib@obs-besancon.fr}
\and 
Department of Physics and Astronomy, University of Nevada, Las Vegas, 4505 South Maryland Parkway, Las Vegas, NV 89154, USA
\and
Aix Marseille Universit\'{e}, CNRS, LAM (Laboratoire d'Astrophysique de Marseille) UMR 7326, 13388, Marseille, France
\and
Center for Radiophysics and Space Research, Space Sciences Building, Cornell University, Ithaca, NY 14853, USA
}

\date{Received ??; accepted ??}

\abstract
{Comets and chondrites show non-monotonic behaviour of their Deuterium to Hydrogen (D/H) ratio as a function of their formation location from the Sun.  This is difficult to explain with a classical protoplanetary disk model that has a decreasing temperature structure  with radius from the Sun.}
{We want to understand if a protoplanetary disc with a dead zone, a region of zero or low turbulence, can explain the measured D/H values in comets and chondrites.}
{We use {time snapshots of a} vertically layered disk model with turbulent surface layers and a dead zone at the midplane. The disc has a non-monotonic temperature structure due to increased heating from self-gravity in the outer parts of the dead zone. We couple this to a D/H ratio evolution model in order to quantify the effect of such thermal profiles on D/H enrichment in the nebula.}
{We find that the local temperature peak in the disk can explain the diversity in the D/H ratios of different chondritic families. 
This disk temperature profile leads to a non-monotonic D/H enrichment evolution, allowing these families to acquire their different D/H values while forming in close proximity. The formation order we infer for these families is compatible with that inferred from their water abundances. However, we find that even for very young disks, the thermal profile reversal is too close to the Sun to be relevant for comets.}
{}
 

\maketitle
\section{Introduction}

The Deuterium-to-Hydrogen (D/H) ratio is one of the most discussed isotopic ratios in planetary sciences. This interest is due to the strong dependence it has on the temperature of an icy body's formation location \citep{review}. In the interstellar medium (ISM), the water D/H ratio is measured to be up to $\sim 9\times 10^{-4}$ \citep{brown}, although the values are much higher in the organic matter of the interstellar hot cores ($\sim 1\times 10^{-3}$) and cold clouds (up to $\sim 2\times 10^{-2}$) \citep{dartois2003,2006mess.book..341R,parise2012}. When a protoplanetary disk forms around a young star, these ices are heated to more than few hundred K. At these high temperatures, gaseous HDO reacts with nebular H$_2$ to form water vapor and HD,
\begin{equation}
HDO + H_2 \rightleftarrows H_2O + HD 
\end{equation}
and this results in a decrease of the D/H ratio in water {because $[HD]/[H_2]$ is lower than D/H in water, even accounting for chemical affinity} \citep{lecluse}. The rate of this reaction depends strongly on the disk temperature. Therefore, ices forming in regions with higher past temperatures should have a lower D/H ratio. D/H measurements in Jupiter and Saturn find D/H $\sim 2 \times 10^{-5}$ \citep{lellouch}, the value also found in the ISM Hydrogen gas. This value is usually interpreted as the D/H ratio in the protosolar nebula's H$_2$. 
For the rest of this work we follow \cite{drouart} and others in defining $f$ as the D/H enrichment factor in water:
\begin{equation}
f=\frac{[HDO]/[H_2O]}{[HD]/[H_2]}
\end{equation}
Thus for the nebular H$_2$ gas, by definition, $f$ is equal to unity. 

\subsection{D/H Ratio Observations in the Solar System}

\begin{table*}
\caption{The measured water D/H fractionation and abundance in small bodies in the solar system.}
\setlength{\tabcolsep}{10pt} 
\renewcommand{\arraystretch}{1.25}
\centering
    \begin{tabular}{  c  c    c}
    \hline
Source & $f$ & \% Water Abundance\\ \hline \hline
   \multicolumn{1}{c}{~~~~~~Ordinary Chondrites (OCs)} \\ \hline 
  LL3 & 29 -- 44 & - \\
  {Semarkona (LL3)} & {$\sim$100} & - \\  
  {H and L} & $\leq$ 15 & 1 \\
  \hline
   \multicolumn{1}{c}{~~~~~~Carbonaceous Chondrites} \\ \hline 
  CI  & 3-5  & 20 \\
  CM  & $\sim$ 4.5   & 13 \\
  CR  & $\sim$ 8 &  6 \\
  CO & 4 -- 7 & 1 \\
  CV & 1 -- 4 & 1 \\
 \hline
  \multicolumn{1}{c}{~~~~~Comets} \\
  \hline
  103P/Hartley & $\sim$ 7.4 &  - \\
   67P & $\sim$ 21 &  \\
   Oort Cloud Comets (OCCs) & 10 -- 25 & -\\
   \hline
   \multicolumn{1}{c}{~~~~~Moons} \\
   \hline 
   Enceladus & $\sim$ 15  &  40-50  \\
    \hline
    \end{tabular}
\end{table*}

The water D/H ratio of small bodies in the solar system has been measured for a large number of comets \citep{mumma}, meteorites \citep{2006mess.book..341R} and in the plumes of Saturn's moon Enceladus \citep{waite, spencer}. In meteorites, both organic materials and water contribute to the bulk D/H value. Therefore, separating the two components is crucial to understanding the contribution from water. {This was done by \cite{alexander}, who found $f$ values ranging from 3 to 5 in carbonaceous CI chondrites, and $\sim$ 15 in ordinary chondrites (OCs). {More recently, a higher value reaching f $\sim$ 100 has been also inferred for OCs \citep{piani2015}}. This implies that either low D/H minerals still coexist in the matrix mixed with deuterium-rich organics, or an unknown process has fractionated water and organics differently from an initial low D/H reservoir, before their incorporation in the matrix. In neither case the low values for the D/H ratio obtained from this method have ever existed in chondrites.} Taken at face value, these results contradict the predictions from the expected formation locations, since CIs are usually associated with C-complex asteroids that formed between the giant planets, and OCs are associated with S-complex asteroids that formed in situ sunward of Jupiter's orbit \citep{s2,as1,walsh}. \cite{alexander2010} noted though that due to Hydrogen escape via Fe oxidation, the D/H ratio in OCs should be treated as an upper value and could have originally been much lower. This effect was found to be limited to OCs, allowing us to use the values measured in other chondritic families to understand the processes at play.

{ The carbonaceous chondrites, CRs, are thought to have a high D/H value \citep{robert1982}.}
The value measured by \cite{alexander} of $f \sim$ 8 is very surprising, since this is almost twice the value in CIs, and is actually higher than that found in the comet 103P/Hartley. The D/H ratio was measured also in the ordinary chondrites type LL3 {and found to be even higher than that in L and H type ordinary chondrites \citep{2006mess.book..341R,alexander2010}}. Measurements of the D/H ratio in the carbonaceous chondrite types CO, CM and CV were found to be also low providing further evidence for this discrepancy in the carbonaceous and ordinary chondrite D/H ratios \citep{2006mess.book..341R,alexander}.

In Oort cloud comets (OCCs), classically thought to have formed between the giant planets, $f$ was found to range between 10 and 25. On the other hand, the D/H ratio was measured for the first time in the Jupiter family comet (JFC) 103P/Hartley, a family classically thought to have formed in the Kuiper belt beyond Neptune, and was found to be $\sim$ 7.4, the value also found in Earth's oceans \citep{hartogh}. This was surprising since JFCs are thought to have formed in an area further out than the OCCs \citep{morby1}  \citep[although cf. the model of][where all comets form beyond the orbital radius of Neptune]{morby2}, their D/H ratio was predicted to be higher than or at least equal to the OCCs \citep{mousis,2011ApJ...734L..30K}. What made the situation even murkier is the recent measurement of the D/H ratio in the JFC 67P/Churyumov–Gerasimenko by Rosetta, where the value found lies well within the OCCs range \citep{altwegg}. A summary of all of the D/H ratio observations discussed here, {and the water abundances in chondrites,} can be found in Table~1.

\subsection{D/H Ratio Connection to Protoplanetary Disc Models}

Classical disk models with a standard monotonically decreasing temperature profile \citep{lyndenbell74,pringle1981} all predict that the D/H ratio should similarly follow a monotonic profile, increasing with the icy body's formation distance from the sun \citep{drouart,mousis,hersant}. An alternative model is that of \cite{yang2013} who use a new D/H 2D model including the infalling material from the cloud, giving a non monotonic D/H ratio profile in the nebula due to constant influx of unequilibrated water. However, they did not discuss the D/H ratio in chondrites, and their monotonic temperature profile out to a radius of $10\,\rm AU$ cannot explain the diversity of D/H ratios found in the inner solar system. Another model is that of \cite{jacquet}, who tried to explain the chondritic diversity with a classical disc model that includes an interplay of inward advection and outward diffusion in the nebula. This model however also predicted a monotonic D/H profile, and although it can broadly explain the chondrites D/H range, it does not explain for example why CRs, that formed closer to the sun than CIs \citep{wood}, have a higher D/H ratio.

The observed D/H fractionation variations found in chondrites are a challenge to classical disk models, since the parent bodies of most chondrites should have formed in the same general region, except those of CIs that probably formed few AU further out in the disk. All these results indicate that either the D/H evolution models used are incomplete or that the thermal profile in the protosolar nebula was not monotonic, the hypothesis we are going to explore in this work.

Turbulence within protoplanetary disks drives outward angular momentum transport that allows material to spiral in and be accreted onto the forming star \citep[e.g.][]{pringle1981}. This turbulence is thought to be driven by the magneto-rotational instability (MRI) \citep{balbus1991}. However, it is now generally accepted that the midplanes of protoplanetary disks have a region of zero or weak turbulence known as a ``dead zone'' \citep[e.g.][]{gammie,fromang,martin2012a,2013ApJ...765..114D,turner,cleeves}. This leads to a gradual accumulation of gas in the dead zone region, resulting in an increase in the temperature and pressure \citep{armitage,zhu,martin2011,martin2013}.  The increased surface density of the disc can lead to a second type of turbulence, driven by gravitational instability \citep{paczynski1978,lodato04} and this results in an increase in the temperature locally in that region. The question we are going to tackle in this work is: \textit{What effect does a local peak in the temperature  have on the D/H ratio in protoplanetary disks and can it resolve the discrepancies in the observations of the D/H ratio in small bodies in our solar system?} In section 2 we discuss the model we use to simulate the processes involved. In section 3 we discuss the results and implications. We draw our conclusions in section 4.

\section{The numerical model}
\subsection{The protoplanetary disk model}
We follow \cite{martin2011} and model the protoplanetary disc as a layered accretion disc (see their equations 1-16). The surface density evolves according to conservation of mass and angular momentum \citep{pringle1981}.  The temperature structure is governed by a simplified energy equation that balances viscous heating with black body cooling (see e.g. \cite{pringle1986,canizzo93}).  There are two types of turbulence in protoplanetary discs, magnetic turbulence driven by the MRI and gravitational turbulence \citep{paczynski1978}. 

The MRI requires a critical level of ionisation to operate so that the gas is well coupled to the magnetic field.  This can be achieved if the temperature of the disc is larger than the critical $T>T_{\rm crit}=800\,\rm K$ \citep{umebayashi}. In this case, the MRI operates at all disc heights. However, the temperature in the outer parts of the disc is lower than this. In this region, the disc surface layers may be ionised by external sources of ionisation such as cosmic rays or X-rays from the central star to a maximum surface density depth of $\Sigma_{\rm crit}=200\,\rm g\,cm^{-2}$ (e.g. \cite{gammie, glassgold04}). If the total surface density of the disc is larger than this, $\Sigma>\Sigma_{\rm crit}$, then there is a dead zone at the midplane with surface density {$\Sigma_{\rm  d}=\Sigma-\Sigma_{\rm crit}$} and the active layers have $\Sigma_{\rm  a}=\Sigma_{\rm crit}$. Otherwise, if $\Sigma<\Sigma_{\rm crit}$, then there is no dead zone layer so that {$\Sigma_{\rm d}=0$} and $\Sigma_{\rm a}=\Sigma$. 

The  MRI active layers have a \cite{shakura} viscosity $\alpha$ parameter of $0.01$ (e.g. \cite{hartmann1998}). In our model, the dead zone has zero turbulence, {unless it becomes self-gravitating}. { Turbulence in the dead zone could be driven by other sources such as hydrodynamic instabilities (including the baroclininc instability, \cite{klahr2003,petersen2007,lesur2010}) or be {induced} from the magnetohydrodynamic instability in the active surface layers \citep{fleming2003,simon2011,gressel2012}.} However, we note that a small amount of turbulence within the dead zone does not significantly alter the qualitative disk structure and behaviour \citep{martin14}. { The conclusions of our work are not significantly affected by an additional source of turbulence within the dead zone, unless the source is strong enough to be able to produce a steady state disk. This would require a level of turbulence comparable to that produced by MRI \citep{martin14}. We discuss the uncertainties associated with the $\alpha$ parameter in Section~3.5.}

Gravitational turbulence requires the \cite{toomre} parameter to be less than the critical, $Q<Q_{\rm crit}=2$. While a dead zone is present in a protoplanetary disc, the flow through the disc is not steady because material accumulates in the dead zone. With sufficient material in the dead zone, it may become self gravitating, thus a small amount of turbulence may be driven. We include an additional viscous term in the surface density evolution equation and a heating term in the temperature equation. This extra heating in the massive dead zone can eventually cause the disc to reach the critical temperature required for the MRI. Once this is reached, there is a snow plough effect through the disc and the whole disc becomes MRI active in an outburst phase \citep{martin2013}. As material drains on the star, the disc cools and the dead zone can reform causing repeating outburst and quiescent cycles. Once the infall accretion rate onto the disc drops off, there may not be sufficient inflow through the disc for another outburst to occur, but there can still be a dead zone within the disc.

{ The infall accretion rate onto a forming star varies in time although the details of the evolution depends on the specific disc model \citep[e.g.][]{Shu1977,Basu1998,Bate2011,Kratter}. However, it is thought that at early times the infall accretion rate is approximately $\dot M_{\rm infall}=c_{\rm s}^3/G$. Thus, assuming a cloud temperature $T\sim 10\,\rm K$, the initial infall accretion rate is $\sim 10^{-5}\,\rm M_\odot\,yr^{-1}$.} In this work, we consider several constant infall rates and thus analyze the disc structure at different evolutionary times. We choose three infall accretion rates onto the disc of $\dot M_{\rm infall}= 2\times 10^{-5}$, $1\times 10^{-6}$ and $1\times 10^{-8}\,\rm M_\odot\,yr^{-1}$.

\begin{figure*}
\begin{center}
\resizebox{\hsize}{!}{\includegraphics[angle=0]{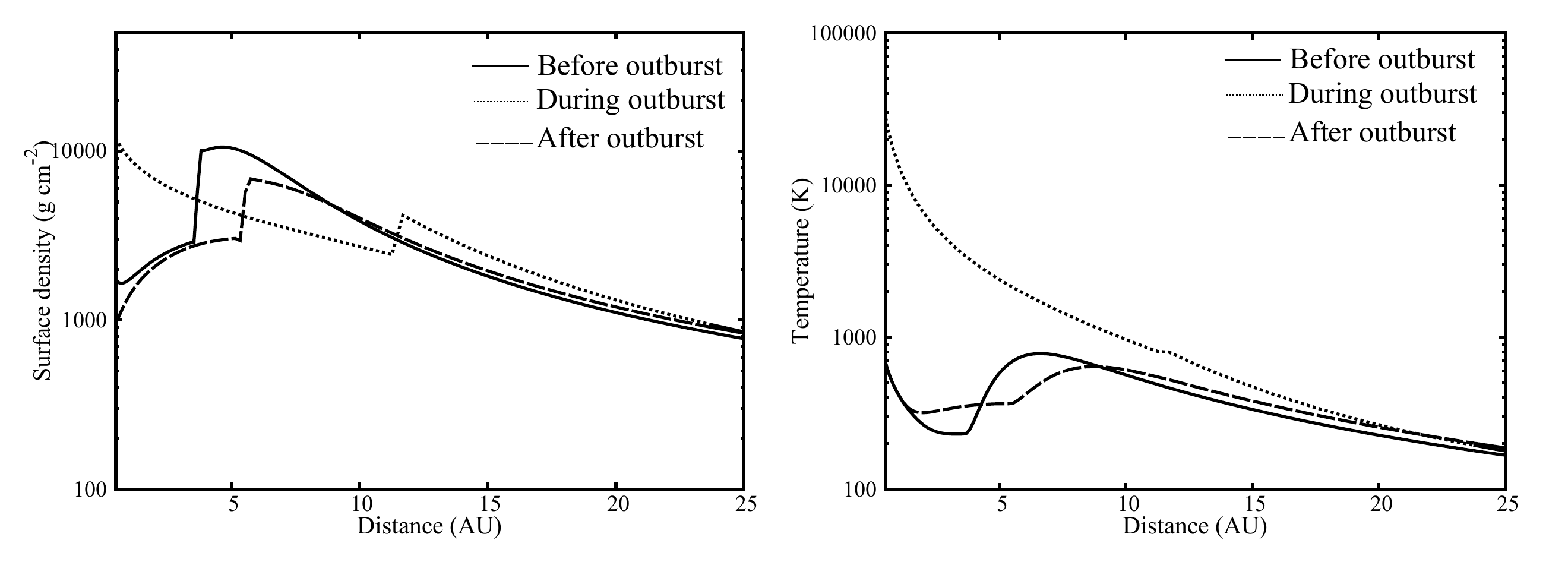}}
\caption{The disk surface density (left panel) and temperature (right panel) structure for the high infall accretion rate $\dot{M}_{\rm infall} = 2\times10^{-5}\,\rm M_\odot yr^{-1}$. The profiles are shown just before, during, and just after an accretion outburst. The ``during outburst'' profile is used in the calculations corresponding to Fig. \ref{duringout}.}
\label{youngdisk}
\end{center}
\end{figure*}

\begin{figure*}
\begin{center}
\resizebox{\hsize}{!}{\includegraphics[angle=0]{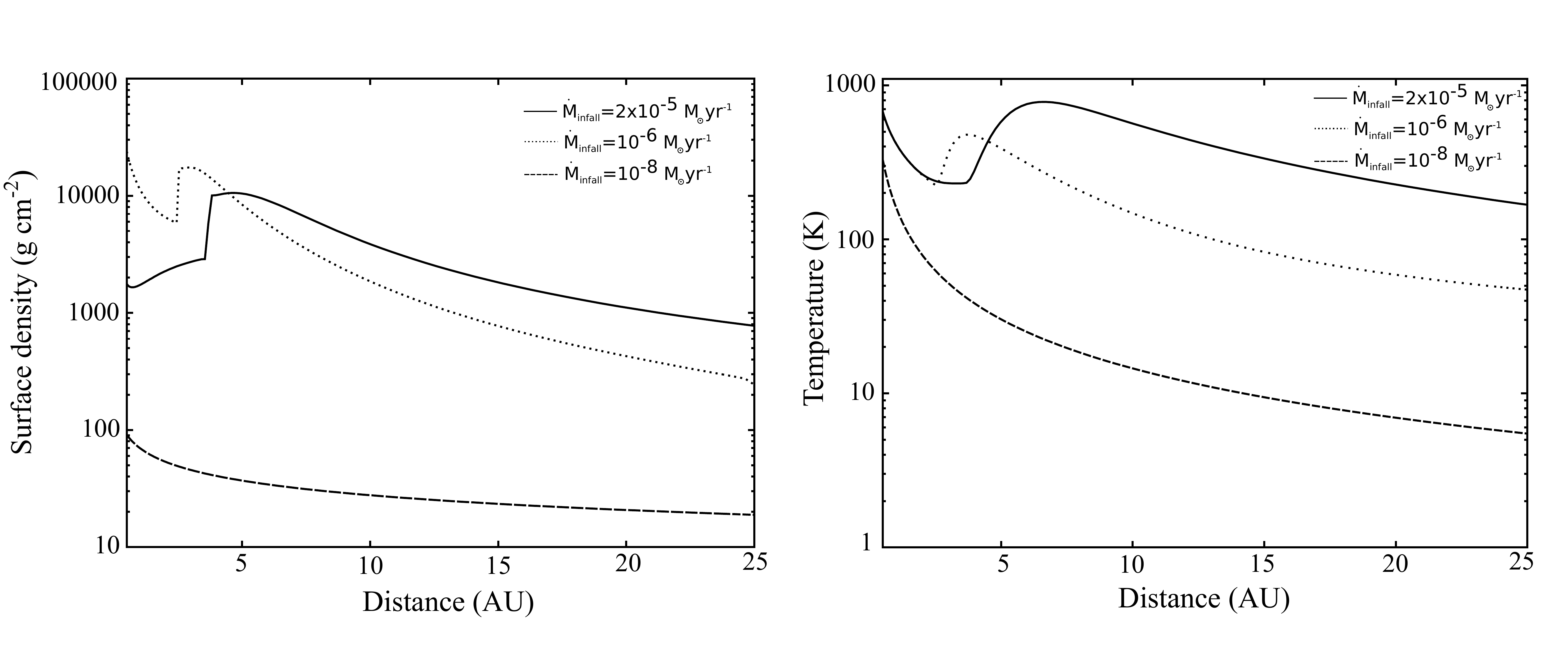}}
\caption{{The disk surface density (left panel) and temperature (right panel) for $\dot{M}_{\rm infall} = 2\times10^{-5} \,\rm M_\odot\,yr^{-1}$ (at a time just before an outburst), $10^{-6} \,\rm M_\odot\,yr^{-1}$ (at a time between outbursts), and $\dot M_{\rm infall}=10^{-8}$ $M_\odot\,yr^{-1}$ (in steady state). The thermal gradient reversal observable in the right panel is main element of the disk model. We use $\dot{M}_{\rm infall} = 1\times10^{-6} \,\rm M_\odot\,yr^{-1}$ as our nominal disk model.}}
\label{disk}
\end{center}
\end{figure*}

We solve the accretion disc equations with 200 grid points spaced equally in $\log R$. For each infall accretion rate, we run the model until the disc reaches either a steady outburst cycle (for the higher two accretion rates), or a steady state (fully MRI active) disc solution (for the lowest accretion rate).  These models represent {different stages of the disc evolution.} In Fig. \ref{youngdisk}  we show the surface density and temperature profiles for the disc as a function of radius at three different times for $\dot{M}_{\rm infall} = 2\times10^{-5}\,\rm M_\odot yr^{-1}$. The first is a time immediately before an outburst, the second is during the outburst, and the third is immediately after the outburst. During this early phase of the disc evolution, the timescale between the outbursts is only $3\times 10^3\,\rm yr$. As we expect the accretion rate on to the disc to decrease exponentially in time, the timescale between the outbursts  increases during the disc lifetime  \citep[see e.g.][]{martin2012b}. Furthermore, the radius of the temperature peak {(the radius at which the disc becomes self--gravitating)} moves inwards in time \citep{martin2013}. In Fig. \ref{disk} we show the surface density and temperature profiles corresponding to the different disk stages and infall accretion rates. For high accretion rates, the local thermal peak is present in the disk, and disappears for $\dot{M}_{\rm infall} \sim 1\times10^{-8}\,\rm M_\odot yr^{-1}$ because there is no dead zone at this low accretion rate. {The model behaves as a classical accretion disk without a dead zone \citep[see for example for comparison][who found temperature and surface density profiles comparable to the model we use for a similar accretion rate value]{baillie2015} }. The position of the thermal peak is also sensitive to the infall accretion rate. The model with $\dot{M}_{\rm infall} = 1\times10^{-6}\,\rm M_\odot yr^{-1}$ is used as the nominal disk model in this work, and this corresponds to the model used in \citep{martin2013}.  We note that there is some uncertainty in the value of $\Sigma_{\rm crit}$ such that it could be much smaller, which would mean that the temperature peak would exist for even lower accretion rates (see section 3.5 for more on this point).

\subsection{The D/H evolution model}
\begin{figure}
\resizebox{\hsize}{!}{\includegraphics[angle=0]{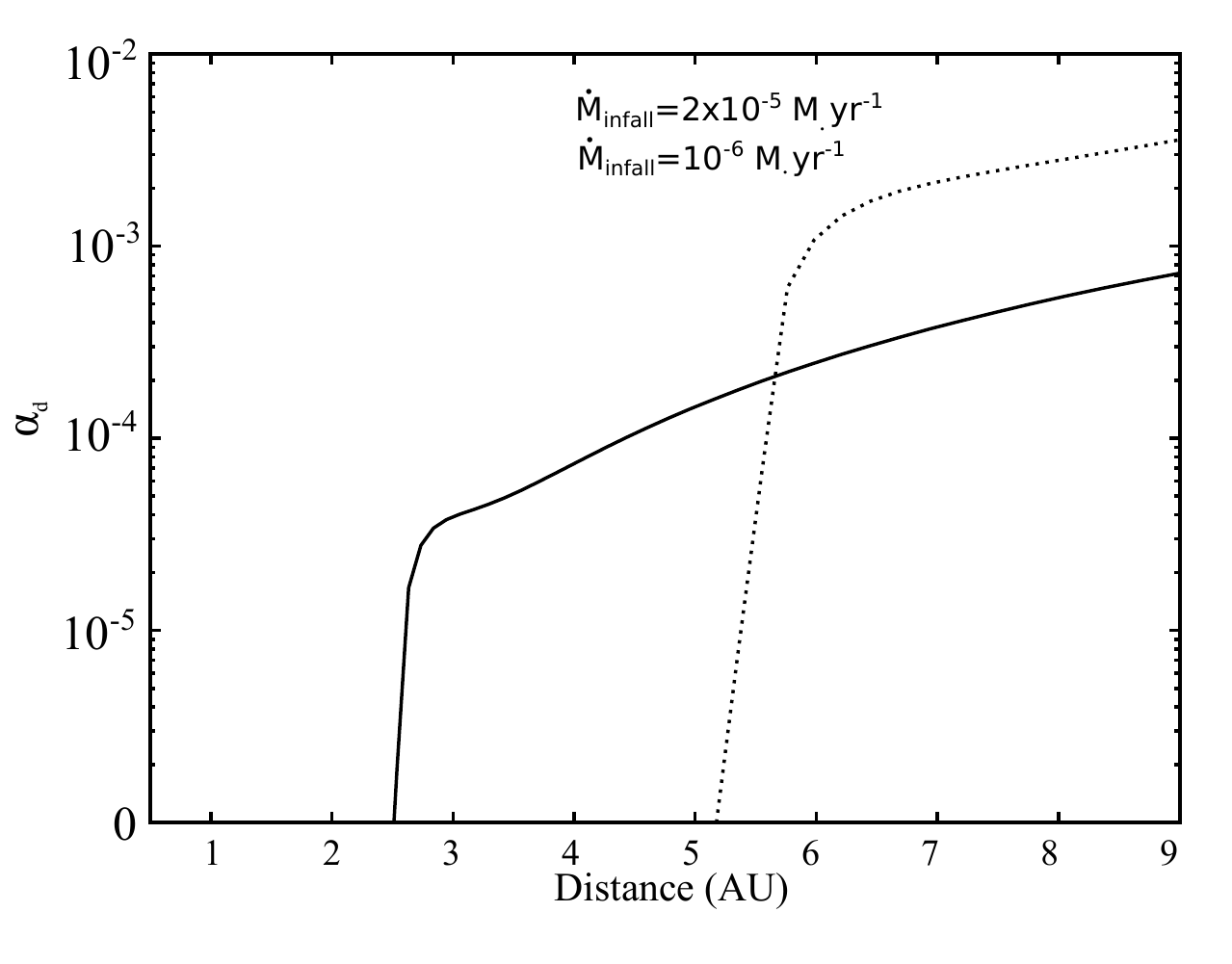}}
\caption{The dead zone midplane turbulent viscosity parameter $\alpha_{\rm d}$. { If the disc is not self--graviting (i.e. $Q>Q_{\rm crit}$), then $\alpha_{\rm d}=0$. Otherwise, its value is depicted by equation~(\ref{ad})}. Note that we manually smooth the viscosity variation (hence $\alpha_{\rm d}$) {over 5 grid points} to avoid code divergence. For $\dot M_{\rm infall}=10^{-6}\,\rm M_\odot \,\rm yr^{-1}$ the smoothing is between 2.2 and 2.5 AU and for $\dot M_{\rm infall}=2\times 10^{-5}\,\rm M_\odot \,\rm yr^{-1}$ this is between 5.2 and 5.5 AU.}
\label{alpha}
\end{figure}

Dust and grains in the protoplanetary disk settle to the disc miplane within the dead zone layer as there is little or no turbulence there.  Thus we expect that the solar systems bodies formed in the dead zone layer. We couple the protoplanetary disk profiles to a classical 1-D D/H ratio evolution model. The D/H ratio  in the dead zone layer is assumed to follow
\begin{eqnarray}
\frac{\partial f}{\partial t} = k P \left(A-f\right)+\frac{1}{\Sigma_d R}\frac{\partial}{\partial R} \bigg(\kappa R \Sigma_d \frac{\partial f}{\partial R} \bigg) 
+ \bigg( \frac{2\kappa}{\Sigma_d} \frac{\partial \Sigma_d}{\partial R} - V_R\bigg) \frac{\partial f}{\partial R}
\end{eqnarray}
\citep{drouart}, where $k(T)$ is the rate of isotopic exchange, $P$ is the gas pressure, $A(T)$ is the fractionation at equilibrium and $\Sigma_{\rm d}$ is the  gas surface density in the dead zone layer. The turbulent diffusivity is
\begin{equation}
\kappa= \frac{\nu_d}{P_r},
\end{equation}
where $\nu_d$ is gas viscosity in the dead zone layer and $P_r$ the Prandtl number. $V_R$ is the radial gas velocity: 
\begin{equation}
V_R=-\frac{3}{\Sigma_{\rm d} R^{1/2}}\frac{\partial}{\partial R}(\nu_{\rm d} \Sigma_{\rm d} R^{1/2})
\end{equation}
\citep{pringle1981}. {This equation is valid in the midplane deadzone layer assuming that $\Sigma_d\gg\Sigma_a$, which is satisfied in the disk profiles that we use (cf. section 3.5 for caveats discussion).} The viscosity due to the self-gravity in the dead zone layer is given by:
\begin{equation}
\nu_{\rm d}=\alpha_{\rm d}\frac{C_{\rm s}^2}{\Omega},
\end{equation}
{where $\alpha_d$ is the turbulence parameter. For all this work, $\alpha_d$, $\Sigma_d$ and $\nu_d$ are the parameters in the midplane (MRI inactive dead zone), and not the value in the MRI active disk surface. $C_s$ is the speed of sound and $\Omega$ the gas Keplerian velocity. The sound speed is calculated with the midplane disc temperature profile.} In this disk model, $\alpha_{\rm d}$ is zero unless the disc locally satisfies $Q<Q_{\rm crit}$, and then we have
\begin{equation}
\alpha_{\rm d}= \begin{cases} \alpha \left[\left(\frac{Q_{\rm crit}}{Q} \right)^2-1  \right] & {\rm if}\,\,\, Q<Q_{\rm crit} \\ 0 &  {\rm otherwise }\\ \end{cases}
\label{ad}
\end{equation}
\citep{martin2011}. { We note that if we were to include an additional source of visosity within the dead zone, other than self--gravity, then we would add an additional term to this equation \citep[see][]{martin14}.}
Fig. \ref{alpha} shows the viscosity parameter in the dead zone for the highest two disk infall accretion rates that we consider. This viscosity in the dead zone is much smaller than that in the active layers. 

The first term on the right side of eq. 3 describes the chemical isotopic exchange between HDO and H$_2$, and takes the pressure and temperature of the disk as the input. The second term describes the gas diffusion due to turbulence that depends on $\alpha$ and $\Sigma$. The third term describes the gas advection that is dependent on $\alpha$, $\Sigma$ and $V_{R}$. In contrast with previous models, we do not neglect this term and we solve the entire equation. The effect of a temperature peak in the nebula in our model is thus controlled by the interplay of these terms. In the absence of turbulence, the first term  dominates and the effect of the temperature peak is to decrease the local D/H ratio. However, if the turbulence is strong enough, the diffusion and advection terms erase this gradient. We run all simulations with $k(T)$ and $A(T)$ from \citep{lecluse}, {and set $P_r$ to 0.5 \citep{pr1,pr2} (although we note that any value of order unity will lead to qualitatively similar results).}

An explicit Forward-time Centered-space (FTCS) scheme was first used to solve this equation. During tests in the case of a regular monotonic thermal profile, the code converged if a sufficiently small time step was used. When the new thermal profiles (with temperature gradient reversals) were used the code diverged even for very small time steps, except for almost vanishing turbulence. Thus, we employ semi-implicit (Crank-Nicholson) and fully implicit schemes to solve the equation over the same grid used in the disk model, {but we take the water snow line (the radial location in the disc inside of which water is gaseous, and outside it is solid, that occurs at a temperature of around $T_{\rm snow}=170\,\rm K$ \citep{lecar} as the outer boundary, since the deuterium exchange can occur only in the vapor phase.}

\section{Results \& discussions}

Most simulations in the literature begin with a spatially constant $f=25$, which is close to the highest value observed today in the solar system {of 29--44} in LL3 meteorites (except Semarkona where a possible value of up to 100 was recently inferred \citep{piani2015}). However, since we are interested in the D/H ratio difference between two bodies rather than the absolute values, we begin our standard simulations with $f_0=15$ which is the average value found in comets. The model evolves from this value and we check the effect of the thermal gradient reversal on the D/H ratio profile. We are therefore implicitly assuming that another transient heating process decreases the initially very high (LL3 or even interstellar) D/H ratios to the lower values we are using as the initial condition. A possible process for this is the gravo-magneto disc instability \citep{martin2011} and its associated accretion outburst that we first test here \citep[see also][]{owen}.  The accretion outburst occurs when the local peak in the temperature profile becomes high enough that it reaches the critical temperature, $T_{\rm crit}$, required to trigger the MRI in the dead zone. During the outburst the disc becomes MRI active throughout and a large amount of material is accreted onto the Sun in a short time. After the outburst, the disk cools, the dead zone reforms and providing that there is sufficient accretion inflow, the cycle repeats.

\begin{figure}
\resizebox{\hsize}{!}{\includegraphics[angle=0]{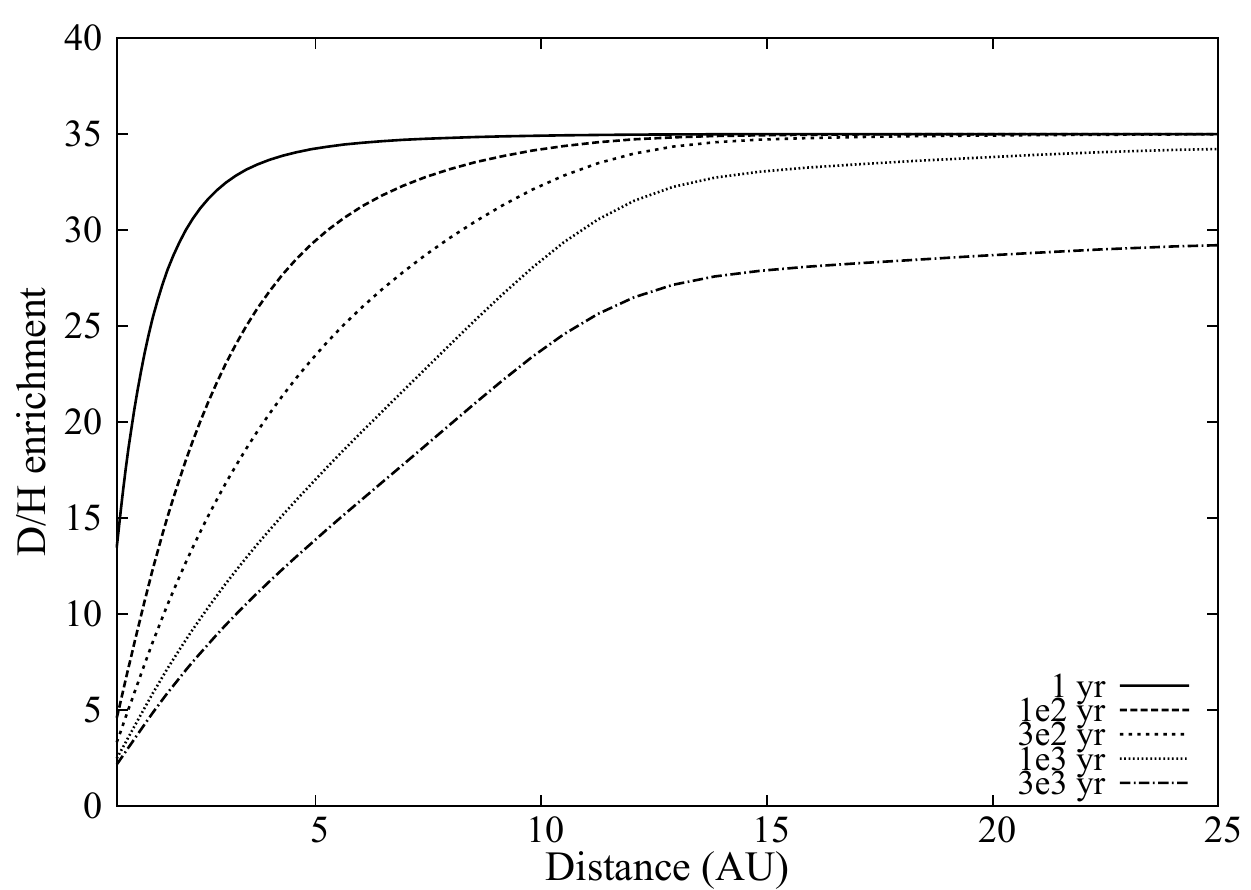}}
\caption{The D/H ratio evolution during an accretion outburst. This simulation starts with $f=35$ as initial condition to reflect the very high enrichment values found often in LL3 and certain ISM regions.}
\label{duringout}
\end{figure}

We first run a test simulation with a disk profile representative of the conditions during an accretion outburst (for $\dot{M}_{\rm infall} \sim 2\times 10^{-5} \,\rm M_\odot yr^{-1}$, cf. Fig. \ref{youngdisk}). We begin with an enrichment factor of $f_0=35$ (representing the very high D/H enrichment that can be found in LL3 or certain ISM environments). The viscosity parameter is $\alpha = 10^{-2}$ everywhere. Further, this leads to fast transport of material. Thus, there is very fast D/H ratio evolution as shown in Fig. \ref{duringout}, where cometary values are reached in the inner 10 AU over the timescale of an outburst. The outburst period lasts a few hundred to few thousand years and during this time the D/H enrichment decreases to cometary or lower values in the inner disk, but remains high in the outer disk. These outbursts can happen multiple times during the disk phase and they alter the D/H ratio inhomogeneities that may have existed prior to the instability. For any process to have measurable effect on the D/H ratio today it has to happen after the last accretion outburst, when the disk infall rate from the cloud has dropped to values below certain threshold point. Thus, in the next Section and for our standard model we consider the evolution at lower infall accretion rates.

\subsection{Standard Model}

We now describe our standard model with an infall accretion rate of $\dot M_{\rm infall}=10^{-6}\,\rm M_\odot\,yr^{-1}$. We consider a time between outbursts as shown in  Fig. \ref{disk}.  We assume that the last accretion outburst happened at an infall accretion rate of around $\dot{M}_{\rm infall}$ reached $1\times 10^{-6}\,\rm M_\odot yr^{-1}$. Thus, the disk structure does not change rapidly after this. The dead zone accumulates material and heats up in the outer parts by self-gravity leading to the thermal gradient reversal, but without reaching sufficiently high temperature ($\sim$ 800 K) to trigger a further outburst. Results for this model are shown in Fig. \ref{main}. We notice that the D/H ratio is decreasing in mainly two locations: the inner hot disk, and around the thermal peak centered at 3.5 AU. The width and limits of the D/H ratio dip around this region are controlled by the turbulence strength. We notice also that in the inner disk, equilibrium is reached in less than 10$^4$ yr, faster than the evolution time of the disk, thus justifying the use of a snapshot disk model in this region. We discuss this further in Section 3.5. Finally, it should be noted that Fig. \ref{main} shows $f$ decreasing to values lower than the average cometary D/H in the outer disk at late times, but this is an artifact of our disk cooling handling as discussed in the next section.

\subsection{Disk cooling and photoevaporation}

As long as there is a hot region {(T higher than $\sim 400 $ K)} along with turbulence within the disk, the global D/H ratio will continue to evolve until it reaches $A(T)$ ($f\sim1$) throughout the disk {(all water equilibrated to nebular gas D/H ratio)}. Such values of $f$ for the entire disk are contradictory to observations. There are two possible explanations for this.

\begin{figure}
\resizebox{\hsize}{!}{\includegraphics[angle=0]{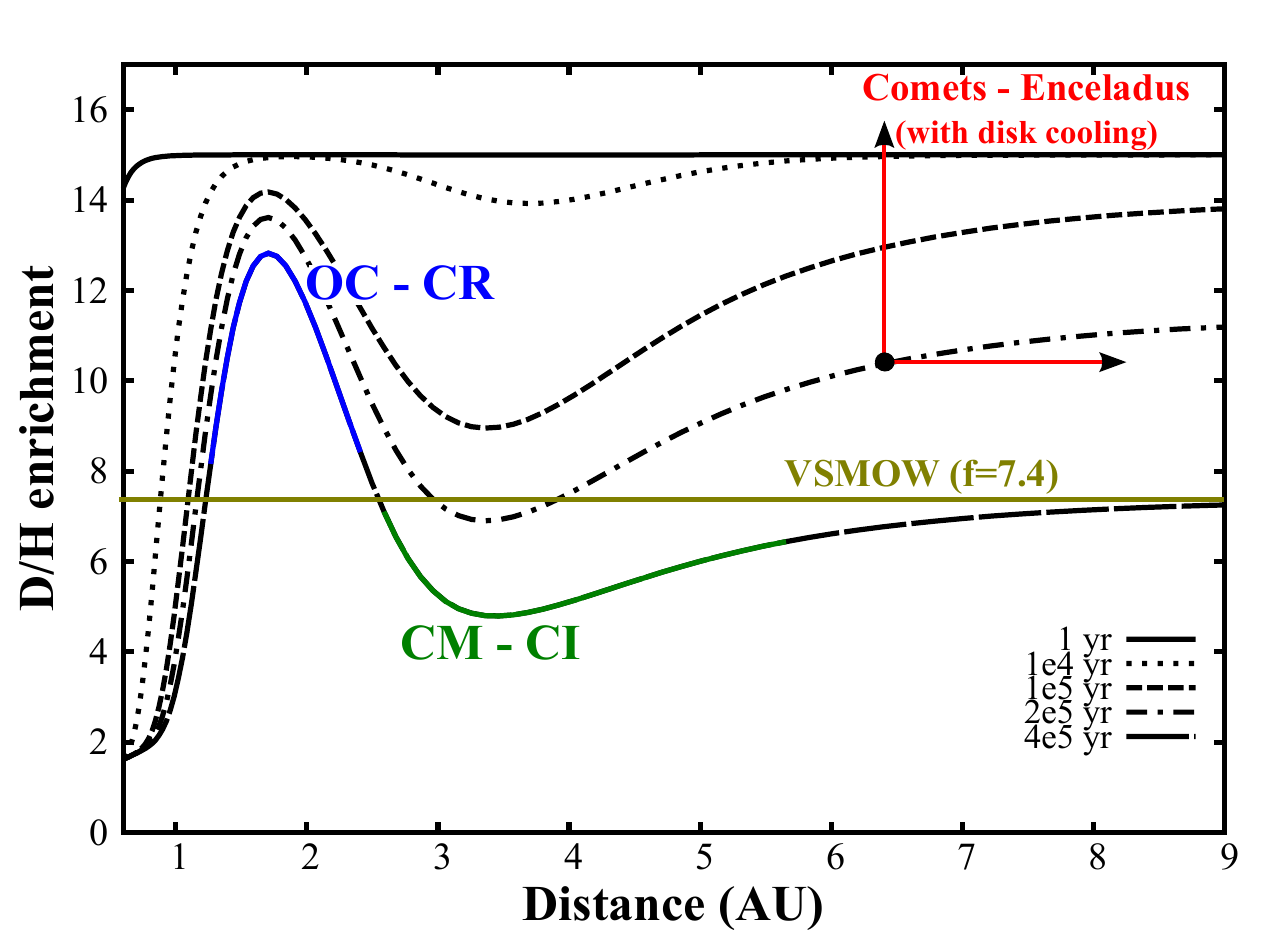}}
\caption{The D/H ratio enrichment evolution for our nominal disk. These profiles allow an important D/H ratio diversity in the chondrites formation region. They explain the values measured in the different chondrites families from their formation location. These inferred formation locations are also compatible with the water abundances of chondrites, with OCs and CRs containing less water (hence probably forming closer to the sun) than CMs and CIs. The exact distance relative to the sun is not important since the model is scalable with $\dot{M}_{\rm infall}$. The D/H ratios in comets, Enceladus, and VSMOW (Earth) are shown for reference. {Having $f$ decreasing to low values in the cometary region at late times is an artifact of our disk cooling handling. It should be higher with consistent cooling treatment. Indeed, when temperature drops below water condensation point, we keep evolving $f$, letting it continue to drop, while it should be fixed at the value reached when the water became isolated from gas by freezing.}}
\label{main}
\end{figure}

First,  the disk could have cooled down sufficiently quickly that the D/H ratio profile became frozen, {as considered in the classical models \citep{mousis,hersant}}. {For low enough temperatures, the chemical exchange (through $k(T)$) becomes very slow and inefficient. The gas-gas reaction then stops completely once water condensed into ice.} This needs to happen before the value of $f$ in the chondrites region becomes too low. This is equivalent to the water snow line radius moving quickly inside the chondrites formation region. The snow line is the radial location in the disk inside of which water is gaseous, and outside it is solid, that occurs at a temperature of around $T_{\rm snow}=170\,\rm K$ \citep{lecar}. For our standard model with an accretion rate of $\dot{M}_{\rm infall}= 1\times 10^{-6} \ M_\odot \ yr^{-1}$, the snow line is at a radius of around $9\,\rm AU$. The disk will remain in this state while $\dot{M}_{\rm infall}$ decreases in time and the snow line moves inwards slowly (\cite{martinlivio2012} and cf. equation 19 in \cite{martin20132}). For the disk parameters we have chosen, our model shows that when $\dot{M}_{\rm infall}$ reaches $\sim 10^{-8} \ M_\odot \ yr^{-1}$, the disk will quickly become cool with a classical monotonic thermal profile with a snow line radius of around $1\,\rm AU$ (see Fig. \ref{disk}, bottom panel). 
The infall accretion rate is given by: 
\begin{equation}
\dot{M}_{\rm infall}=\dot{M}_{\rm i} \exp\bigg(-\frac{t}{t_{\rm ff}}\bigg)
\end{equation}   
\citep[equation 19 in][]{martin2012b}, where $\dot{M}_{\rm i}$ is the initial infall accretion rate (in this case we take $\dot M_{\rm i}=10^{-6} \ M_\odot \ yr^{-1}$), $t$ is time, $t_{\rm ff}$ is the free fall time scale \citep[we take $t_{\rm ff}=10^5 \,yr$ e.g.][]{armitage}). This equation shows that $\dot{M}_{\rm infall}$ will reach $\sim 10^{-8} \ M_\odot \ yr^{-1}$ in about 5$\times10^5$ yr. We will hence stop our simulations at this time, and use the end state results to fit the measurements. We are therefore implicitly assuming that the transition to a cold classical MRI active disk happens quickly, with the snowline moving in from the giant planets region to around 1 AU, thus freezing the D/H ratio in the chondrites regions. The final D/H profile in Fig. \ref{main} is thus in the solid phase.

The second explanation for how the D/H radio profile became frozen is that the disk was completely photoevaporated on a similar timescale to the evolution of $f$. During the photoevaportation process at the end of the disk lifetime the disk is dispersed on a short timescale of around $10^5\,\rm yr$ {\citep{photoev1,photoev3,photoev2}}. Thus, the profile for the ratio of D/H would become fixed in the chrondrites at this time and similarly, the shape of the profile shown in Fig.~\ref{disk} would become fixed at this time.

\begin{figure*}
\begin{center}
\label{nosl}
\end{center}
\end{figure*}

\subsection{Implications for chondrites}

Now we discuss the implications of our model for chondrites. {Figure \ref{main} shows our nominal D/H profile obtained for our nominal model, with some known chondritic D/H ratio ranges.} The profile shows a clear peak reaching $f \sim 15$ around 2 AU, followed by a dip reaching $f \sim 4$ around 3.5 AU. The curve then increases all the way to the cometary values. The absolute peak and dip positions with respect to the sun in this plot are not too important, since these are sensitive to $\dot{M}_{\rm infall}$, and thus can be varied (since the chondrites parent bodies probably formed slightly outward of these positions, to allow for S-type asteroids to form around 2.2 AU). We chose a single model for simplicity, and this particular value to remain consistent with \cite{martin2013} to allow for comparison. As seen in Figure \ref{main}, this profile can fit the D/H values found in most chondrites, and explain their diversity from their formation location.  Thus, this explains why CRs have a higher D/H value than other type of chondrites. The relative formation locations in this plot are also compatible with their water abundances \citep{wood,Brearley}, {although this abundance alone has its limitations due to the very nature of chondrites as an association of different components. Other indicators such as chondrules abundance should be taken into account}. CIs and CMs are more aqueously altered than CRs and OCs that are more reduced. Assuming naively that water abundance increases with heliocentric distance (although cf. the dynamical water distribution models of for example \cite{ciesla,ali-dib}), this implies that CIs and CMs formed further out than CRs and OCs, which is compatible with this profile, and as proposed by \cite{wood}. We can tentatively try to fit the remaining chondrite families as well (COs and CVs) that are more reduced than CMs but less than OCs. The D/H profile does contain regions that fit their values, however we note that there are many uncertainties present in the model since it is only a proof of concept. A caveat in this model is that CIs form too close to CMs, although a distance of several AU might be needed between the formation locations of the two to allow the formation of chondrules.
\subsection{Implications for comets}

The next step is to check if such model can explain the D/H ratio diversity in comets. Naively one can expect that earlier in the disk lifetime, when $\dot{M}_{\rm infall}$ was higher, the thermal peak could have existed further out in the disk, maybe in the comets region. This can lead to a D/H profile analogous to Fig. \ref{main}, but further out in disk, with its own peak and dip. This dip can give a neat explanation for the relatively low D/H value in 103P/Hartley, with the other comets forming in other locations. Assuming a classical comets formation model with the JFCs forming further out than OCCs, 103P/Hartley could have formed on the D/H dip, while 67P/C-G formed slightly further out outside of the dip, and the OCCs forming inside of it. To test this hypothesis we used a profile derived from our disk simulation with $\dot{M}_{\rm infall}=2\times 10^{-5}\ M_\odot \ yr^{-1}$ corresponding to the time just before an outburst. Results are shown in Fig. \ref{disk} (top panel). At this stage of the disk evolution, the thermal reversal (and the corresponding D/H enrichment dip) are around 8 AU, further out than in our canonical case, and as expected from an early disk. The thermal reversal's position, even for such young disk, is still too close in to be relevant for comets formation. Another problem posed by this profile is that it almost certainly leads to an outburst, homogenizing the D/H ratio across the disk. Within the model and parameters we use, we are unable to explain the low D/H ratio in 67P/C-G. The possibility that there exists another set of parameters and/or assumptions within the same framework that can lead to a thermal reversal in the comets region is not excluded though, and is left to future work. {We note that all cometary D/H models assume that the value measured in the comet's ejecta reflect its bulk value, although \cite{podolak2002} showed that the nuclei D/H ratio might be different that on the surface. Additionally, experiments by \cite{brown2} showed that the measured value might also change a function of the instrument-target distance.} 

Another seemingly unrelated problem that can be addressed using such models is the origin of crystal silicates and CAIs in comets. These minerals can form only at temperatures in the order of thousands of kelvins, but they are found in comets \citep{campins,wooden,chi,kelley}. How the high temperature minerals got to the cold region where comets form is a classic problem. {Some of the proposed solutions were outward turbulent diffusion of particles \citep{morvan} and photophoresis \citep{mousis2007}}. The recent observation of narrow crystal silicates features in the spectra of a young solar like star during an accretion outburst indicated that these outbursts might be the formation mechanism of high temperature minerals \citep{abraham}. {For the accretion outburst  for $\dot{M}_{\rm infall} \sim 2\times 10^{-5}\,\rm M_\odot yr^{-1}$, the outburst trigger radius (the radius of the temperature peak) in our model is around 7 AU  (cf. Fig. \ref{youngdisk}), considerably widening high temperature minerals formation region. {It should be mentioned that most materials inside of the trigger radius get accreted onto the sun during the accretion outburst, so only elements forming beyond this radius remain in the disk.} Quantifying any of these possible solutions is beyond the scope of this work. }

\subsection{Caveats}
Since this work was only intended to be a proof of principle highlighting the concept and quantifying the relative strengths of diffusion and chemistry in a non monotonic nebula, numerous assumptions and simplifications were made in this model:
\begin{itemize}
\item Ideally, one should use a time evolving disk model coupled dynamically with the D/H module, to track the simultaneous evolution of both components. However, for simplicity we used a static (snapshot) disk profile with the time dependent D/H module. Hence we are making the implicit assumption that the D/H ratio evolves on a shorter timescale than the disk. This assumption is justified by the short timescale of the D/H evolution ($10^{4}-10^{5}$ yr) compared to the disk evolution timescale ($\sim 10^6\,\rm yr$).
\item Our disk profiles are derived from a layered (active and dead zones) disk. In this work, we are only tracking the D/H ratio in the midplane (dead zone) and ignoring any effect the active layer might have, including the sedimentation of equilibrated water. Our work is thus valid only if the dead zone surface density is much higher than the active layer surface density. Since our domain starts beyond the dead zone inner boundary at $0.5\,\rm AU$ and extends out to the snow line radius at $9\,\rm AU$ {(for the particular choice of the infall rate and thus disk age)}, much closer than the dead zone outer boundary at $23\,\rm AU$. This validity condition for our model is thus applicable throughout the entire domain. 
\item We note that there are several unknown parameters in the layered disc model. For example, the critical surface density that is ionised by external sources is not well determined. Dead zone models that include more physics generally find active layer surface densities that may be very small (e.g. \cite{bai2011}). However, such small active layers cannot explain accretion rates observed in T Tauri stars (e.g. \cite{2011ApJ...727....2P,martin2012b}). Thus we fold all of the uncertainty into the parameter $\Sigma_{\rm crit}$. The value of $\alpha$ in the active layers is also not well determined. However, these parameters do not affect the qualitative behavior of the disc.
\item {In this work we started our main simulation (Fig. \ref{main}) from a constant D/H value throughout the disk. Realistically however the preceding outburst may lead to a heterogeneous D/H distribution. Quantifying this effect needs a fully time dependent coupled disk-D/H evolution, for a smooth temperature variation to occur. This is left for future work.}  
\item {Recently, 2D (r-z) steady--state models of protoplanetary discs have been constructed with {an $\alpha$-variation} over the disc height to mimic the effects of a reduced (but non--zero) $\alpha$ in a dead zone \citep{bitsch2014}. These disc models do not find the increase in temperature in the dead zone region present in our disc models because self--gravity does not operate. There is not a sufficient build up of material in their disc models to cause the disc to become self--gravitating. This is because the chosen values for $\alpha$ are high { enough that} a steady state disc is found. \cite{martin14} showed that even with some turbulence in the dead zone, the qualitative disc behavior is as we have described in this paper, unless the turbulence in the dead zone is comparable to that in the active layer, where a steady state may be found. Previous 2D (r-z) simulations that are time--dependent and included a dead zone with a smaller viscosity agree with the numerical models used in this work \citep{zhu2009}. Further detailed { magnetohydrodynamic} time-dependent numerical simulations are required in order to determine the correct value of $\alpha$ in the dead zone {\citep[see for example][]{Simon2013}}.}
\item {In our simplified model, we set the viscosity in the dead zone to be zero, except where it is generated by self--gravity. It is possible for other hydrodynamical instabilities to operate in the dead zone (for example the baroclinic instability, vertical shear instability and others \citep{turner}. However, as discussed by \cite{martin14}, the qualitative disc behavior is the same even if there is a small amount of turbulence in the dead zone. The temperature peak is still there, and the outbursts still occur.} 
\end{itemize}

\section{Conclusions}
We have coupled a D/H enrichment code including diffusion, advection and chemical exchange {to snapshots from protoplanetary disk model} that includes a dead zone. The disk model contains a local temperature peak at a radius of around 3 AU due to the heating by self-gravity in the outer parts of the dead zone. 
We found that 
this leads to a dip in the D/H profile around the same region, in contrast with the classical monotonic D/H models. The new profile can explain the origin of the D/H ratio variations between the different chondrites families. We propose that CI chondrites (that have a relatively low D/H ratio) formed in the region of the thermal gradient reversal, but CRs (that have a high D/H ratio) formed just inside of this region. The new D/H profile also accommodates the formation of COs, CVs, and CMs. However, even with a younger disk profile the model is unable to explain the the D/H ratio in 67P/C-G.  The thermal gradient reversal is too close to the Sun to be relevant. Finally we proposed that the accretion outbursts associated to these models can explain the presence of high temperature minerals across the disk.  

{This work shows that detailed temperature profiles from {time-dependent layered} disk models provide a potential explanation for the rich variation of D/H ratios found in the solar system.} A more detailed understanding of the role the thermal inversions in dead zones and outbursts plays in shaping the chemistry of the nebula necessitate a more elaborate exploration of the parameters space, to be the subject of future works.

\begin{acknowledgements}
Special thanks go to D. Bockel\'{e}e-Morvan and E. Lellouch for useful discussions on comets. We thank the two anonymous referees for useful comments. M.A.-D was supported by a grant from the city of Besan\c{c}on.
O.M. acknowledges support from CNES. This work has been partly carried out thanks to the support of the A*MIDEX project (n\textsuperscript{o} ANR-11-IDEX-0001-02) funded by the ``Investissements d'Avenir'' French Government program, managed by the French National Research Agency (ANR). JIL acknowledges support from the JWST program through a grant from NASA Goddard. 
\end{acknowledgements}

\end{document}